\documentclass[twocolumn,showpacs,preprintnumbers,amsmath,amssymb]{revtex4}

\usepackage{graphicx}
\usepackage{bm}

\newcommand{\epsfigwide}[5]{%
\begin{figure*} \vspace{#3}%
\includegraphics{#2}%
\caption{
\label{fig:#1} #5}
\vspace{#4}
\end{figure*}}

\newcommand{\tsub}[1]{_{\mbox{\scriptsize#1}}}

\newcommand{\quarterthin}{\kern 0.0417em}
\newcommand{\comm}[2]{[ \quarterthin #1 , #2 \quarterthin ]}
\newcommand{\bra}[1]{\langle#1|}
\newcommand{\ket}[1]{|#1\rangle}
\newcommand{\ev}[1]{\langle#1\rangle}

\begin{document}
\draft



\title{
SO(5) as a Critical Dynamical Symmetry \\
in the SU(4) Model of High-Temperature Superconductivity }

\author{Lian-Ao Wu$^{(1,2)}$, Mike Guidry$^{(3)}$,
Yang Sun$^{(3) \footnote{Present address: 
Department of Physics, University of Notre Dame, Notre Dame, IN 
46556}}$,
and Cheng-Li Wu$^{(4)}$
}
\address{
$^{(1)}$Department of Physics, Jilin University, Changchun, Jilin, 
130023, PRC\\
$^{(2)}$Department of Chemistry, University of Toronto, Ontario M5S 
3H6, Canada\\
$^{(3)}$Department of Physics and Astronomy, University of 
Tennessee, Knoxville, Tennessee 37996 \\
$^{(4)}$Physics Division, National Center for Theoretical Science, 
Hsinchu, Taiwan 300, ROC}

\begin{abstract} 
An $SU(4)$ model of high-temperature superconductivity and
antiferromagnetism has recently been proposed. The $SO(5)$ group employed by 
Zhang is embedded in this 
$SU(4)$ as a subgroup, suggesting a connection between our $SU(4)$ model
and the Zhang $SO(5)$ model. In order 
to understand the relationship between the
the two models, we have used generalized coherent states to
analyze the nature of the $SO(5)$ subgroup.  By constructing coherent-state 
energy
surfaces, we demonstrate explicitly that the $SU(4)\supset SO(5)$ symmetry can 
be interpreted as a critical dynamical symmetry interpolating between 
superconducting
and antiferromagnetic phases, and that this critical dynamical symmetry has many
similarities to critical dynamical symmetries identified previously in
other fields of physics. More generally, we demonstrate with this example that  
the mathematical techniques associated with generalized coherent states may have
powerful applications in condensed matter physics because they provide a clear
connection between microscopic many-body theories and their broken-symmetry
approximate solutions. In addition, these methods may be interpreted
as defining the most general Bogoliubov transformation subject to a Lie group
symmetry constraint, thus providing a mathematical connection between algebraic
formulations and the language of quasiparticle theory.  Finally, we suggest that
the identification of the $SO(5)$ symmetry as a critical dynamical symmetry 
implies deep algebraic connections between high-temperature superconductors and 
seemingly unrelated phenomena in other field of physics.
\end{abstract}

\date{\today}
\maketitle



\section{Introduction}

Data for cuprate high-temperature superconductivity (SC) suggests
$d$-wave singlet pairing in the superconducting state  and that 
superconductivity 
in these systems is closely related to the antiferromagnetic (AF) insulator
properties of the undoped compounds.  This proximity of superconducting and
antiferromagnetic order is unusual and suggests that a correct description of
cuprate superconductors must permit the SC and AF order to enter the theory on a
similar footing.

\subsection{Unified Models}

S.-C. Zhang et al
\cite{zha97,hen98,rab98} proposed to unify AF and SC states by assembling their
order parameters into a 5-dimensional vector and constructing an $SO(5)$ group 
that
rotates AF order into $d$-wave SC order. Recently, we introduced 
an $SU(4)$ model \cite{gui99} 
of high-temperature SC and AF order having $SO(5)$ as a subgroup. 
The motivation for and methodology of the $SU(4)$ model are superficially rather
different from those of Zhang and collaborators.  However, the appearance of the
Zhang $SO(5)$ as a subgroup of the $SU(4)$ symmetry implies that the two models
must have a strong physical relationship, as we have already suggested
\cite{gui99}.  

One difficulty in
understanding this relationship is that the two models are formulated using
different approaches that employ different languages. In the $SU(4)$ model the
five operators responsible for AF order (three staggered magnetization operator
components) and  SC order (two $d$-wave pair operators) enter the theory as
quantum
mechanical operators, on exactly the same footing as all of the other ten 
generators
of $SU(4)$ \cite{note1}.  Thus, the $SU(4)$ model is a many-body, fully quantum
mechanical theory in which the charge and spin are exactly conserved, so there
is no spontaneously broken symmetry.  This is, of course, as it should be: the
charge and spin are rigorously conserved in a full many-body formulation of the
problem as they are in nature.
In the Zhang $SO(5)$ model, the five operators corresponding to the staggered
magnetization and the $d$-wave singlet pairing are instead treated as order
parameters (expectation values of operators in a broken symmetry state).
Therefore, in the $SO(5)$ model the antiferromagnetic phase and the
superconducting phase are associated with {\em approximate solutions} of the
many-body problem that break charge and spin symmetry spontaneously.

The methodology employed in the development of the $SU(4)$ model (systematic
application of principles of dynamical symmetry) has found broad application in
other fields such as nuclear structure \cite{wu94,IBM}, molecular physics
\cite{Ia95,Ia99}, or particle physics \cite{Bi94}.  
This implies a deep algebraic connection between high temperature
superconductivity and a variety of phenomena in other fields of physics that 
bear no superficial resemblance to high-temperature superconductors.  
For example, we have pointed
out previously \cite{gui99}
that there is an almost perfect mathematical analogy at the algebraic level
between the AF-SC competition in cuprates and the competition 
between quadrupole deformation of the nuclear mean field and nucleon pairing 
that is a central organizing principle of nuclear structure physics.  
The appearance of the $SO(5)$ subgroup in the dynamical symmetry of the
$SU(4)$ model for high-temperature superconductors 
is very similar to that of the $SO(7)$ subgroup in the $SO(8)$
fermion dynamical symmetry model \cite{wu94} that describes nuclear
structure.  

\subsection{Generalized Coherent States}

There is a well-developed theoretical approach to  relating a many-body 
algebraic
theory with no broken symmetry to an approximation of that theory that exhibits
spontaneously broken symmetry: the method of generalized coherent states
\cite{wmzha90}. This method may be viewed as the extension of Glauber coherent
state theory (which is built on an $SU(2)$ Lie algebra) to a more complex system
having an arbitrary Lie algebra structure. It has also been shown to be
equivalent to the most general Hatree-Fock-Bogoliubov variational method
under symmetry constraints, and has been applied extensively in various areas of
physics and mathematical physics, though not to our knowledge in condensed 
matter.  

The result of such a coherent state
analysis is a set of energy surfaces that represent an approximation to the
original theory in which order parameters appear as independent variables. In 
the
general case, these energy surfaces can exhibit (possibly multiple) minima and
these minima  may appear at non-zero values of the order parameters, implying
spontaneous symmetry breaking. Thus, the generalized coherent state method is a 
systematic approach to relating a many-body algebraic theory to its approximate
symmetric and broken symmetry solutions.  

\subsection{Critical Dynamical Symmetries}

The concept of a {\em critical dynamical symmetry} appears naturally in
applications of generalized coherent state techniques  
to other fields of physics. 
A critical dynamical symmetry 
is a dynamical symmetry having eigenstates that vary
smoothly with a parameter (usually particle-number related) such that the
eigenstates approximate one phase of the theory on one end of the parameter 
range
and a different phase of the theory at the other end of the parameter range, 
with
eigenstates in between exhibiting large softness against fluctuations in the 
order
parameters describing the two phases.  

We shall demonstrate here what is, to our knowledge, the first example in 
condensed
matter physics of such a symmetry.  In this case, the critical dynamical 
symmetry
will be shown to be based on the $SO(5)$ subgroup of $SU(4)$, and it will be 
shown
to interpolate between AF and SC order as the hole doping parameter is varied. 
Thus, we shall propose that, within the context of the more general $SU(4)$ 
model,
the $SO(5)$ symmetry employed by Zhang 
serves as a doorway between AF and SC order in a
manner that can be specified in precise terms using the language of Lie algebras
and generalized coherent state theory, and that is related to critical dynamical
symmetries that have been found in other fields of physics. 

\subsection{Symmetry Breaking}

We shall use these results to derive a result that has received
considerable attention for the $SO(5)$ model:  that an exact $SO(5)$ symmetry
cannot account for the detailed phenomenology of the cuprate superconductors and
that it is necessary to break $SO(5)$ (explicitly, not spontaneously) in a
particular way in order to recover Mott insulator normal states 
at half filling, 
as is required by the data. As we shall show, the embedding of $SO(5)$ as a 
subgroup
of $SU(4)$ implies naturally that $SO(5)$ must be broken in this manner in 
order
to produce the correct normal states at half filling.

\subsection{Goals of Paper}

Let us conclude this introduction by enumerating concisely the primary goals of
this paper:  

\begin{enumerate}

\item This paper serves to introduce the generalized coherent state method to
issues in condensed matter physics. 

\item We shall show how to relate the generalized coherent state to the most
general variational quasiparticle states that can be constructed subject to the 
constraints of $SU(4)$ symmetry.

\item This paper introduces into the high-temperature superconductor discussion 
in particular and condensed matter in general, 
the concepts of critical dynamical
symmetries that have been applied with considerable success in other fields of
physics. 

\item We shall demonstrate that the coherent state solution of the
$SU(4)$ model identifies $SO(5)$ as a critical dynamical symmetry. 
This critical dynamical symmetry will be shown to
interpolate between AF and SC order as the hole doping of the system is varied. 
This doping dependence is a natural consequence of the $SU(4)$ symmetry, without
the introduction of a chemical potential {\em ansatz}.

\item We shall show that 
the AF and SC phases themselves
are more economically described, not by $SO(5)$, but by dynamical symmetries 
built
on $SO(4)$ and $SU(2)$ subgroups of $SU(4)$, respectively. Thus, we shall
demonstrate that $SU(4)$ accounts for both the origin of the AF and SC order
parameters and the $SO(5)$ ``rotation'' of the superspin vector between these
phases.

\item We shall demonstrate that as a fundamental consequence of the $SU(4)$ 
structure the
$SO(5)$ subgroup {\em must be broken} in order to produce Mott insulator normal
states.  Furthermore, we shall demonstrate that the required symmetry breaking
terms and the doping dependence of the solutions occur naturally within the 
$SU(4)$
parent algebra and need not be introduced by hand as is required in the Zhang $SO(5)$
model.

\item We shall show that, because of the nature of the critical dynamical
symmetry,
an AF perturbed $SU(4)\supset SO(5)$ Hamiltonian is able to approximate various
symmetry limit solutions depending on the doping: the solutions are close to the
$SO(4)$ limit presenting AF order around half filling, and approach the
$SU(2)$ limit  presenting SC order as the hole doping increases. Thus the
$SU(4)$
model with a perturbed $SO(5)$ Hamiltonian is able to account for the essential
features of high $T_c$ superconductors.

\end{enumerate}

Our approach will be to introduce the basic features of the coherent state 
technique in sections II and III. In section IV we derive the coherent state 
energy
surfaces for the $SU(4)$ model, and in Section V we discuss the $SU(4)$ energy
surfaces in various dynamical symmetry limits.  In Section VI, we use
$SU(4)$ energy surfaces to examine the properties of broken $SO(5)$ symmetry, 
and
then use these results in Section VII to argue that with a small AF-preferred
symmetry breaking, an $SU(4)\supset SO(5)$ 
Hamiltonian may be able to describe high
temperature superconductivity. Section VIII presents a summary and conclusions.

\section{Coherent States and the SU(4) Matrix Representation}

Gilmore \cite{gil72,gil74} and Perelomov \cite{per72} (see also earlier work by
Klauder \cite{kla63}) demonstrated that Glauber coherent states
\cite{gla63} for the electromagnetic field could be generalized to define 
coherent
states associated with the structure of an arbitrary Lie group.  In particular,
they observed that the original Glauber theory 
for coherent photon states may be 
expressed in terms of an $SU(2)$ Lie algebra by examining the commutation
properties of the second-quantized operators of the theory.  Once the theory has
been formulated in terms of an $SU(2)$ algebra generated by combinations of
creation and annihilation operators, the formalism may be generalized to 
encompass a set of such operators closed under any Lie algebra.   

We shall term this extension of the Glauber theory to arbitrary Lie algebras 
the
{\em generalized coherent state method.}  Since an extensive review concerning 
the basic approach has been presented in \cite{wmzha90}, 
we omit an introduction 
to the general technique and proceed directly to a specific application of the
generalized coherent state method to the $SU(4)$ algebra.

\subsection{Algebraic Structure}

A convenient way to analyze these generalized coherent states is in terms of 
their
geometry, which is in one-to-one correspondence with the coset space. Let us 
begin
with the algebraic structure of the $SU(4)$ model \cite{gui99}. We introduce 16 bilinear
fermion operators:
\begin{eqnarray} 
p_{12}^\dagger&=&\sum_k g(k) c_{k\uparrow}^\dagger
c_{-k\downarrow}^\dagger
\qquad p_{12}=\sum_k g^*(k) c_{-k\downarrow} c_{k\uparrow} \nonumber
\\ q_{ij}^\dagger &=& \sum_k g(k) c_{k+Q,i}^\dagger c_{-k,j}^\dagger
\qquad q_{ij} = (q_{ij}^\dagger)^\dagger \label{eq1}
\\ Q_{ij} &=& \sum_k c_{k+Q,i}^\dagger c_{k,j} \qquad S_{ij} = \sum_k
c_{k,i}^\dagger c_{k,j} - \tfrac12 \Omega \delta_{ij}  \nonumber
\end{eqnarray} 
where $c_{k,i}^\dagger$ creates a fermion of momentum $k$ and 
spin projection $i,j= 1 {\rm\ or\ }2 \equiv \ \uparrow$ or
$\downarrow$, $Q=(\pi,\pi,\pi)$ is an AF ordering vector, $\Omega/2$ is the
electron-pair degeneracy, and following Refs.\ \cite{hen98,rab98} we define 
\begin{displaymath}
g(k) = {\rm sgn} (\cos k_x -\cos k_y)
\end{displaymath} 
with the constraints
\begin{displaymath} 
g(k+Q) = -g(k), \qquad
\left| g(k) \right| = 1.
\end{displaymath} 
Under commutation the operator set (\ref{eq1}) closes a
$U(4)$ algebra corresponding to the group structure
\begin{eqnarray} 
&\supset& SO(4) \times U(1) \supset SU(2)\tsub{s} \times U(1)
\nonumber
\\ U(4) \supset SU(4) &\supset& SO(5) \supset SU(2)\tsub{s} \times U(1)
\label{eq2}
\\ &\supset& SU(2)\tsub{p}
\times SU(2)\tsub{s} \supset SU(2)\tsub{s} \times U(1) 
\nonumber
\end{eqnarray} 
where we require each subgroup chain to end in the subgroup
\begin{equation} SU(2)\tsub{s} \times U(1)
\label{conserved}
\end{equation}
representing spin (the $SU(2)\tsub{s}$ factor) and charge (the $U(1)$ factor)
conservation, because the physical states of the system obey these conservation
laws.

In Ref.\ \cite{gui99} we discussed the representation structure of (\ref{eq2}) and
showed that the $SO(4)$ subgroup is associated with antiferromagnetism, the
$SU(2)\tsub{p}$ subgroup is associated with
$d$-wave superconductivity, and the $SO(5)$ subgroup is associated with a
transitional symmetry interpolating between the other two. In this paper, we
further provide the full mathematical justification for interpreting the $SO(5)$ 
subgroup
as a symmetry interpolating dynamically between SC and AF phases.

\subsection{A Convenient Basis for Generators}

It is convenient to take as the generators of $U(4) \rightarrow U(1)_{\rm cd}
\times SU(4)$ the new combinations
\begin{eqnarray} 
Q_+&=&Q_{11}+Q_{22} = \sum_k (c_{k+Q\uparrow}^\dagger
c_{k\uparrow} + c_{k+Q\downarrow}^\dagger c_{k\downarrow}) \nonumber
\\
\vec S &=& \left( \frac{S_{12}+S_{21}}{2},
                \ -i \, \frac {S_{12}-S_{21}}{2},
                \ \frac {S_{11}-S_{22}}{2} \right) \nonumber
\\
\vec {\cal Q} &=& \left(\frac{Q_{12}+Q_{21}}{2},\ -i\, \frac{Q_{12}-Q_{21}}{2},
\ \frac{Q_{11}-Q_{22}}{2} \right)
\label{eq3}
\\
\vec \pi^\dagger &=& \left( i\frac {q_{11}^\dagger - q_{22}^\dagger}2, \
\frac{q_{11}^\dagger + q_{22}^\dagger}2,
\ -i\frac {q_{12}^\dagger + q_{21}^\dagger}2 \right) \nonumber
\\
\vec \pi&=&(\vec \pi^\dagger)^\dagger
\quad D^\dagger = p^\dagger_{12}
\quad D = p_{12}
\quad M=\tfrac12 (S_{11}+S_{22}) \nonumber
\end{eqnarray} 
where $Q_+$ generates the $U(1)_{\rm cd}$ factor and is 
associated
with charge density waves (do not confuse this $U(1)$ factor with the one 
appearing
in Eq.\ (\ref{conserved}) that is associated with charge conservation),
$\vec S$ is the spin operator, $\vec {\cal Q}$ is the staggered magnetization, 
the operators $\vec \pi^\dagger$ and $\vec \pi$ are those of Ref. \cite{zha97}, 
the operators $D^\dagger$ and $D$ are associated with
$d$-wave pairs, and 
\begin{displaymath} 
2M=n-\Omega
\end{displaymath}
is the charge operator. Because of the direct product structure 
$U(4) \rightarrow U(1)_{\rm cd} \times SU(4)$, we can without loss of generality
analyze the $U(4)$ structure in terms of its subgroup $SU(4)$, with the 
$U(1)_{\rm
cd}$ factor considered separately.  Hence all subsequent discussion will deal 
with the $SU(4)$ subgroup of $U(4)$.

To facilitate comparison with the $SO(5)$ symmetry, the
$SO(6) \sim SU(4)$ generators may be expressed as
\begin{equation} L_{ab} =
\left(
\begin{array}{cccccc} 0&&&&&
\\ D_+&0&&&&
\\
\pi_{x+} & -{\cal Q}_x & 0 &&&
\\
\pi_{y+} & -{\cal Q}_y & -S_z & 0 &&
\\
\pi_{z+} & -{\cal Q}_z & S_y & -S_x & 0 &
\\ i D_- & M & i\pi_{x-} & i \pi_{y-} & i\pi_{z-} & 0
\end{array}
\right )
\label{eq4}
\end{equation} 
where we define
\begin{equation} 
D_\pm = \tfrac12 (D \pm D^\dagger)
\qquad
\pi_{i\pm} = \tfrac12 (\pi_i \pm  \pi^\dagger_i)
\label{eq5}
\end{equation} 
with $L_{ab} = -L_{ba}$ and with commutation relations
\begin{equation}
\comm{L_{ab}}{L_{cd}} = i(\delta_{ac} L_{bd} -\delta_{ad}L_{bc} -\delta_{bc}
L_{ad}+ \delta_{bd}L_{ac}).
\label{eq6}
\end{equation}

\subsection{Faithful Matrix Representation}

The coherent state method requires a faithful matrix representation of
$SU(4)$.  Explicit multiplication verifies that the following mapping preserves
the algebra of (\ref{eq6}):
\begin{equation} 
\begin{array}{llll} p_{12}^{\dagger }\rightarrow & \left[
\begin{array}{cc}\ \,0 & \hspace{11pt}i\sigma _y\hspace{5.5pt} \\ \ \,0 &0
\end{array}
\right] & p_{12}^{}\rightarrow& \left[ \begin{array}{cc} 0 & \hspace{9pt}0 \\
-i\sigma _y & \hspace{9pt}0
\end{array}\hspace{4pt}
\right] \vspace{8pt} \\ q_{12}^{\dagger }\rightarrow & \left[
\begin{array}{cc}\ \,0 & \hspace{11pt}\sigma _x\hspace{8.5pt} \\ \ \,0 &
0\end{array}
\right]& q_{12}^{}\rightarrow&  \left[ \begin{array}{cc} 0 & \hspace{9pt} 0 \\
\hspace{5pt} \sigma _x \hspace{5pt} &\hspace{9pt}0
\end{array}\hspace{5pt}
\right]\vspace{8pt}  \\ q_{11}^{\dagger }\rightarrow & \left[
\begin{array}{cc}\ \,0 &\hspace{2pt}I+\sigma _z \\ \ \,0 & 0
\end{array}
\right] & q_{11}^{}\rightarrow& \left[ \begin{array}{cc} 0 &
\hspace{2pt}0\hspace{2pt} \\ I+\sigma _z & \hspace{2pt} 0\hspace{2pt}
\end{array}\
\right]\vspace{8pt}  \\ q_{22}^{\dagger }\rightarrow & \left[
\begin{array}{cc} 
\ \, 0 & \hspace{2.5pt} I-\sigma _z \\ \ \,0 & 0
\end{array}
\right] & q_{22}^{}\rightarrow& \left[ 
\begin{array}{cc} 0 & \hspace{2pt}0 \\ I-\sigma _z & \hspace{2pt}0
 \end{array}\hspace{5.5pt}\right]\vspace{8pt}  \\ S_{12}^{}\rightarrow & \left[
\begin{array}{cc}
\ \sigma _{+} & \ \ 0 \\ 0 & \hspace{4pt} -\sigma _{-}
\end{array}
\right]   & S_{21}^{}\rightarrow& \left[ \begin{array}{cc}
\hspace{5pt} \sigma _{-} \hspace{5pt}& 0 \\ 0 & \hspace{-2.5pt}-\sigma _{+}
\end{array}
\right] \vspace{8pt}   \\ S_{11}^{}\rightarrow & \left[
\begin{array}{cc}
\frac{I+\sigma _z}2 & 0 \\ 0 & \hspace{-10pt}-\frac{I+\sigma _z}2
\end{array}
\right]   & S_{22}^{}\rightarrow& \left[ \begin{array}{cc}
\frac{I-\sigma _z}2 & 0 \\ 0 & \hspace{-10.5pt}-\frac{I-\sigma _z}2
\end{array}
\right] \vspace{8pt}  \\ Q_{12}^{}\rightarrow & \left[
\begin{array}{cc}
\ \sigma _{+} &\hspace{4pt}\ \,0 \\ 0 & \hspace{9pt}\sigma _{-}
\end{array}\
\right] & Q_{21}^{}\rightarrow& \left[ \begin{array}{cc}
\ \sigma _{-} & \hspace{10pt}0 \\ 0 & \hspace{9pt}\sigma _{+}
\end{array}\
\right]\vspace{8pt} \\
\tilde{Q}_{11}^{}\rightarrow & \left[ \begin{array}{cc}
\frac{I+\sigma _z}2 & 0 \\ 0 &\hspace{-2pt} \frac{I+\sigma _z}2
\end{array}
\right]  & \tilde{Q}_{22}^{}\rightarrow& \left[ \begin{array}{cc}
\frac{I-\sigma _z}2 & 0 \\ 0 &\hspace{-2.5pt} \frac{I-\sigma _z}2
\end{array}
\right] 
\end{array}
\label{eq7}
 \end{equation} 
where 
\begin{displaymath}
\tilde{Q}_{ii}\equiv Q_{ii}+\tfrac{\Omega}{2},
\end{displaymath} 
the $\sigma _x$, $\sigma _y$, and $\sigma _z$ are Pauli matrices 
in the standard representation,
\begin{displaymath}
\sigma _{\pm } \equiv \tfrac 12(\sigma _x\pm i\sigma _y),
\end{displaymath}
and $I$ is a unit matrix.  Likewise, it is easily verified that in terms of 
this representation,
\begin{equation}
\begin{array}{ll}
\hspace{-1pt}D^{\dagger}\hspace{-1pt} \rightarrow \hspace{-1pt}\left[
\begin{array}{cc} 0 & i\sigma _y \\ 0 &
 0 \end{array}\right]  &\hspace{24pt} D\rightarrow \left[ 
\begin{array}{cc} 0 & 0 \\ \hspace{-2pt}-i\sigma _y & 0\end{array}\right]
\vspace{10pt}
\\
\pi^{\dagger}_x\rightarrow \left[
\begin{array}{cc} 0 & i\sigma _z \\ 0 & 0\end{array}\right] & \hspace{24pt}
\pi^{ }_x\rightarrow \left[
\begin{array}{cc} 0 & 0 \\ \hspace{-4pt}-i\sigma _z & 0\end{array}\right]   
\vspace{10pt}
\\
\pi^{\dagger}_y\rightarrow\left[
\begin{array}{cc} 0 & \hspace{5pt}I \hspace{4pt} \\ 0 &\hspace{4pt} 0
\hspace{4pt}\end{array}\right]   & \hspace{24pt}
\pi^{ }_y\rightarrow\left[
\begin{array}{cc}\hspace{4pt} 0 & \hspace{5pt}0 \\ \hspace{6pt} I \hspace{2pt}&
\hspace{5pt}0\end{array}\right] 
\vspace{10pt}
\\
\pi^{\dagger}_z\rightarrow \left[
\begin{array}{cc} 0 & \hspace{4pt}\sigma _x \\ 0 & 0\end{array}\right]  &
\hspace{24pt}
\pi^{ }_z\rightarrow \left[
\begin{array}{cc} \hspace{4pt}0 &\hspace{3pt} 0 \\ \hspace{5pt}\sigma _x &
\hspace{3pt}0\end{array}\right] 
\vspace{10pt}
\\\vec{Q}\hspace{3pt}\rightarrow \left[
\begin{array}{cc}\vec{S} & \hspace{2pt}0\hspace{4pt} \\ 0 &
\hspace{2pt}\vec{S}\hspace{4pt}\end{array}\right] & \hspace{24pt}
\hspace{2pt}\vec{S}\hspace{2pt}\rightarrow \left[
\begin{array}{cc}\hspace{4pt}\vec{S} & \hspace{8pt}0 \\ \hspace{4pt}0
&\hspace{-1.5pt} -\vec{S}\end{array}\right]
\vspace{10pt}
\\ M\rightarrow \left[
\begin{array}{cc} \frac{I}{2} & 0\\ 0 &\hspace{-1pt} -
\frac{I}{2}\end{array}\right]  & \hspace{24pt}
\tilde{Q}_{+}\hspace{-4pt}\rightarrow \left[ 
\begin{array}{cc}\hspace{4pt} I & \hspace{4pt}0 \\ \hspace{4pt}0 \hspace{4pt}&
\hspace{4pt}I\hspace{2pt}\end{array}\right]
\end{array}\label{SD}
\end{equation} 
where $\tilde{Q}_{+}={Q}_{+}+\Omega$.

\subsection{Collective Subspace}

We take as a Hilbert space 
\begin{displaymath}
| {\cal S} \rangle = \ket{n_x n_y n_z n_d} =
(\pi_x^\dagger)^{n_x} (\pi_y^\dagger)^{n_y} (\pi_z^\dagger)^{n_z} 
(D^\dagger)^{n_d}
\ket{0}
\end{displaymath} 
which is a collective subspace associated with $SO(6)$ irreps of 
the form $$ (\sigma_1,\sigma_2,\sigma_3) = (\tfrac \Omega2,0,0).
$$ In this notation we use the well-known isomorphism of $SU(4)$ and $SO(6)$ to
label the irreducible representations with the standard $SO(6)$ quantum numbers
$(\sigma_1, \sigma_2, \sigma_3)$
\cite{su4so6}.   Physically, this irrep represents a ``maximally stretched'' 
state
in the representation space that is in turn associated with maximal 
collectivity;
as such, it is the obvious candidate  for a collective subspace describing the
lowest states of the system.

One sees immediately that the expectation value of $Q_+$ is zero for any state 
in
this collective $SU(4)$ space: the matrix representation of $\tilde{Q}_+$ is a 
unit matrix, and thus $Q_+$ commutes with all the $SU(4)$ generators, leading to
\begin{displaymath}
\bra{{\cal S}}Q_+\ket{{\cal S}}=0.
\end{displaymath}
In the symmetry limit,  this implies that charge density wave excitation are
excluded from the 
$SU(4)$ model restricted to this subspace \cite{CDwave}.   

\subsection{SU(4) Casimir Operator}

The Casimir operator of
$SU(4) \sim SO(6)$
\begin{equation} 
C_{su(4)}=\vec \pi^\dagger \hspace{-2pt}\cdot \hspace{-2pt}\vec
\pi + D^\dagger  D +
\vec S \cdot \vec S + \vec {\cal Q}
\cdot\vec {\cal Q} + M(M-4) 
\label{eq8}
\end{equation} 
is an invariant and its expectation value in this collective subspace  
\begin{equation}
\bra{{\cal S}} C_{su(4)}\ket{{\cal S}} = \tfrac{\Omega}{2}(\tfrac{\Omega}{2}+4)
\label{casimir}
\end{equation}  
is a constant.

\section{Symmetry-Constrained Bogoliubov Transformation}

Utilizing the methods of Ref.\ \cite{wmzha90}, the coset space is 
$$ SU(4)/SO(4)\times U(1),
$$ where the
$SO(4)$ subgroup is generated by $\vec{\cal Q}$ and
$\vec S$, and $U(1)$ is generated by the charge operator $M$. The coherent state
may be written as
\begin{equation}
\mid \psi\rangle={\cal T}\mid 0^* \rangle.
\label{eq10}
\end{equation} 
The operator ${\cal T}$ is defined by
\begin{equation} 
{\cal T} = \exp (\eta_{00} p_{12}^{\dagger }+\eta_{10}
q_{12}^{\dagger }-{\rm h.\ c.}), 
\label{eq10b}
\end{equation} 
where $|0^{*}\rangle$ is the physical vacuum (the ground state of
the system), the real parameters $\eta_{00}$ and $\eta_{10}$ are
symmetry-constrained variational parameters, and h.\  c.\ means the hermitian
conjugate. Since the variational parameters weight the elementary excitation
operators
$p^\dagger_{12}$ and $q^\dagger_{12}$ in Eq.\ (\ref{eq10b}), they represent
collective state parameters for a subspace truncated under the $SU(4)$ symmetry.
The most general coherent state corresponds to a 4-dimensional, complex, compact
manifold parameterized by 8 real variables. The reduction of the coherent state
parameters to only two in Eq.\ (\ref{eq10b}) follows from requiring time 
reversal symmetry and assuming the conservation of spin projection $S_z$ 
for the wavefunction.

It is often simpler to view the coherent states as Hatree-Fock-Bogoliubov (HFB)
variational states constrained by the dynamical symmetry. The symmetry-
constrained
HFB coherent state method is discussed in 
Refs. \cite{wmzha90,wmzha87,wmzha88a,wmzha88,wmzha89}. It may be viewed as a 
type of
mean-field approximation to the underlying many-body problem that is 
particularly
useful in the present  context because it leads to easily visualized energy
surfaces. This identification provides a natural connection to spontaneously 
broken
symmetries and effective Lagrangian field theories on the one hand, and to
quasiparticle language on the other. 

From the coset representative expressed in the 4-dimensional matrix 
representation
(\ref{eq7}), the transformation operator 
${\cal T}$ defined in Eq.\ (\ref{eq10}) may be written as  
\begin{displaymath} 
{\cal T} = \left[
\begin{array}{cc} {\bf Y}_1 & {\bf X} \\ -{\bf X}^{\dagger } & {\bf Y}_2
\end{array}
\right]
\quad {\bf X} \equiv \left[
\begin{array}{cc} 0 & \alpha + \beta \\ - (\alpha - \beta) & 0
\end{array}
\right]
\end{displaymath} 
where ${\bf Y}_1$ and ${\bf Y}_2$ are determined by the 
requirement
that ${\cal T}$ be unitary, and $\alpha$ and $\beta$ are variational parameters 
related to $\eta_{00}$ and $\eta_{10}$  in Eq.\ (\ref{eq10b}) (see Ref.\
\cite{wmzha90}). Introducing 
\begin{equation}
\mbox{v}_+ \equiv \alpha+\beta\hspace{24pt}\mbox{v}_- \equiv
\alpha-\beta ,
\label{abuv}
\end{equation} 
the requirement of unitarity gives
$$ {\bf X} = \left[
\begin{array}{cc} 0 & \mbox{v}_+\\ -\mbox{v}_- & 0
\end{array}
\right]
\quad {\bf Y}_1 = \left[
\begin{array}{cc}
\mbox{u}_+ & 0\\ 0 & \mbox{u}_-
\end{array}
\right]
\quad {\bf Y}_2 = \left[
\begin{array}{cc}
\mbox{u}_- & 0\\ 0 & \mbox{u}_+
\end{array}
\right]
\label{XY-matrix}
$$ 
with the constraint that
\begin{equation}
\mbox{u}_\pm^2+\mbox{v}_\pm^2=1 
\label{uveq1}
\end{equation}

\subsection{Quasi-fermion Transformation}

The existence of the 4-dimensional matrix representation of the $SU(4)$ algebra
implies the existence of a representation 
in which the single-particle basis can 
be written in the form 
$$
\{c^\dagger_{\sigma\uparrow},c^\dagger_{\sigma\downarrow},
c_{\bar{\sigma}\uparrow},c_{\bar{\sigma}\downarrow} \},
$$
where
$\bar{\sigma}$ is a state conjugate to $\sigma$. In this representation, any
generator $\hat{O}$  may be written  as
\begin{eqnarray} 
\left[
\begin{array}{cc} O^{11} &  O^{12} \\  O^{21} &  O^{22}
\end{array}
\right]&\rightarrow& \hat O=  
\sum_{\sigma,i,j} \left[ O^{(11)}_{ij}c^\dagger_{\sigma i} c_{\sigma j}
+O^{(22)}_{ij}c_{\bar{\sigma} i} c^\dagger_{\bar{\sigma} j}\right.\nonumber\\  
&+&
\left.O^{(12)}_{ij} c^\dagger_{\sigma i} c^\dagger_{\bar{\sigma} j} +
O^{(21)}_{ij} c_{\bar{\sigma} i} c_{\sigma j }\right ],
\nonumber 
\end{eqnarray} and Eq.\ (\ref{eq10}) is seen to be a Bogoliubov type
transformation, but one that is constrained to preserve the $SU(4)$ symmetry.  

Through the operator
$\cal T$, the physical vacuum state
$|0^*\rangle$  (the ground state of the system) is transformed to a 
quasiparticle
vacuum state
$|\psi\rangle$, with the parameters $\alpha$ and $\beta$ (or $\mbox{v}_\pm$) 
determined by minimizing the energy of the system. Likewise, the basic fermion
operators 
$$
\{c^\dagger_{\sigma\uparrow},c^\dagger_{\sigma\downarrow},
c_{\bar{\sigma}\uparrow},c_{\bar{\sigma}\downarrow} \}
$$
 are transformed to quasifermion operators
$$
\{a^\dagger_{\sigma\uparrow}, a^\dagger_{\sigma\downarrow},
a_{\bar{\sigma}\uparrow}, a_{\bar{\sigma}\downarrow} \}
$$ through
\begin{equation} {\cal T}
\left (
\begin{array}{c} c_{{\sigma}\uparrow}
\\ c_{{\sigma}\downarrow}
\\ c^\dagger_{\bar\sigma\uparrow}
\\ c^\dagger_{\bar\sigma\downarrow}
\end{array}
\right) = 
\left(
\begin{array}{c} a_{{\sigma}\uparrow}
\\ a_{{\sigma}\downarrow}
\\ a^\dagger_{\bar\sigma\uparrow}
\\ a^\dagger_{\bar\sigma\downarrow}
\end{array}
\right).
\label{transf}
\end{equation}

\subsection{One-Body and Two-Body Operators}

Using the transformation (\ref{transf}), one can express any one-body operator 
in the quasiparticle space as \cite{ring80}  
\begin{eqnarray} 
{\cal T}\hat O{\cal T}^{-1}
       &=&\left[\begin{array}{cc} {\cal O}^{(11)} & {\cal O}^{(12)} \\   {\cal
O}^{(21)} & {\cal O}^{(22)}\end{array}\right]\nonumber\\
\rightarrow \hat{\cal O}&=& \sum_{\sigma,i} {\cal O}^{(22)}_{ii}\nonumber
 +\sum_{\sigma,i,j} \left\{ {\cal O}^{(11)}_{ij}a^\dagger_{\sigma i} a_{\sigma 
j} -
{\cal O}^{(22)}_{ji} a^\dagger_{\bar{\sigma} i} a_{\bar{\sigma}
j}\right.\nonumber\\ &+& \left.{\cal O}^{(12)}_{i,j} a^\dagger_{\sigma i}
a^\dagger_{\bar{\sigma} j} + {\cal O}^{(21)}_{i,j} a_{\bar{\sigma} i} a_{\sigma 
j}
\right\}
\nonumber 
\end{eqnarray} 
where the ${\cal O}^{(\mu\nu)}_{ij}$'s are fixed by the
transformation properties of the operator $\hat O$:
\begin{equation} {\cal O}^{(\mu\nu)}_{ij} =\sum_{m,n} [\ {\cal T}^{(\mu 
m)}O^{(mn)}
({\cal T}^{-1})^{(n \nu)} ]_{ij}
\label{omn}
\end{equation}  
and ${\cal T}^{(\mu m)}$ and $O^{(mn)}$ are  two-dimensional 
submatrixes of ${\cal T}$ and $\hat O$, respectively.

Because the quasiparticle annihilation operator acting on the quasiparticle 
vacuum
$|\psi\rangle$ is zero, the expectation values for one-body operators 
$\hat{O}$ are given by
\begin{equation}
\langle\hat O\rangle = \langle\psi|\hat{\cal O}|\psi\rangle =
\sum_{\sigma,i} {\cal O}^{(22)}_{ii}=\sum_\sigma{\rm Tr} ({\cal O}^{(22)})
\label{onebody}
\end{equation} and for two-body operators $\hat O_A \hat O_B$, 
\begin{eqnarray}
\langle\hat O_A \hat O_B\rangle &=& 
\langle\psi|\hat{\cal O_A}\hat{\cal O_B}|\psi\rangle \nonumber\\ &=&
\sum_\sigma{\rm Tr} ({\cal O}^{(22)}_A) \sum_{\sigma'}{\rm Tr} ({\cal
O}^{(22)}_B)\label{twobody}
\\ &+&\sum_\sigma {\rm Tr} ({\cal O}^{(21)}_A {\cal
O}^{(12)}_B). \nonumber
\end{eqnarray}

\subsection{Expectation Value of Generators}

Utilizing equations (\ref{SD}) and (\ref{omn})--(\ref{twobody}), and noting
that the summation $\sum_\sigma$ provides a factor of $\Omega/2$ since the  
matrix elements of Eq.\ (\ref{omn}) do not depend on $\sigma$, one obtains the
expectation values for all the generators and their scalar products in the 
coherent state representation: 
\begin{eqnarray}
\langle D^{\dagger}\rangle &=&\langle D\rangle=\tfrac12 \Omega
(\mbox{u}_-\mbox{v}_- +\mbox{u}_+\mbox{v}_+),
\label{eqd}
\\\langle \pi^{\dagger}_z\rangle &=&\langle \pi_z\rangle=\tfrac12 \Omega
(\mbox{u}_-\mbox{v}_- -\mbox{u}_+\mbox{v}_+),
\label{eq13a2}
\\ Q  &\equiv& \langle {\cal Q}_z\rangle  =
\tfrac12 \Omega (\mbox{v}_+^2 - \mbox{v}_-^2), 
\label{eq14}
\\
\langle\hat n\rangle &\equiv&\ \ n\ \,=\Omega (\mbox{v}_+^2 +
\mbox{v}_-^2),
\label{eqn}
\\
\langle \pi_x\rangle &=& \langle
\pi_y\rangle=\langle \vec{S}\rangle = \langle {\cal Q}_x\rangle=\langle {\cal
Q}_y\rangle=0 ,
\\
\langle D^{\dagger }D\rangle &=&\tfrac14 \Omega^2 (\mbox{u}_+\mbox{v}_+ +
\mbox{u}_-\mbox{v}_-)^2 +
\tfrac12 \Omega (\mbox{v}_+^4 + \mbox{v}_-^4), \hspace{20pt}
\label{eqd2}
\\
\langle \overrightarrow{\pi}^{\dagger
}\hspace{-2pt}\cdot\hspace{-2pt}\overrightarrow{\pi}\rangle &=&\tfrac14 \Omega^2
(\mbox{u}_+\mbox{v}_+ -
\mbox{u}_-\mbox{v}_-)^2 \nonumber \\
& &+
\tfrac12 \Omega (\mbox{v}_+^4 + \mbox{v}_-^4+4\mbox{v}_+^2 \mbox{v}_-^2),
\\
\langle \overrightarrow{{\cal Q}}
\cdot \overrightarrow{{\cal Q}}\rangle &=&\tfrac14 \Omega^2 (\mbox{v}_+^2 -
\mbox{v}_-^2)^2+ \tfrac12 \Omega [(\mbox{u}_+\mbox{v}_+)^2 
\label{eqq2}
\nonumber
\\ & &+ (\mbox{u}_-\mbox{v}_-)^2 +(\mbox{u}_+\mbox{v}_-)^2
+(\mbox{u}_-\mbox{v}_+)^2\,],
\\
\langle \overrightarrow{S}
\cdot \overrightarrow S\rangle &=& \tfrac12 \Omega [(\mbox{u}_+\mbox{v}_-)^2 +
(\mbox{u}_-\mbox{v}_+)^2\,],
\\
\langle M^2\rangle &=& \tfrac14 (n-\Omega)^2+\tfrac12 \Omega
[(\mbox{u}_+\mbox{v}_+)^2 + (\mbox{u}_-\mbox{v}_-)^2\,] .
\end{eqnarray} 
Using the above results, one can also verify Eq.\ (\ref{casimir}) 
explicitly for the expectation value of the Casimir operator $C_{su(4)}$.

\subsection{Order Parameters}

By virtue of the unitarity condition (\ref{uveq1}), there are only two  
independent
variational parameters in the above equations.  They may be chosen as either
$\mbox{v}_+$ and $\mbox{v}_-$, or as $\alpha$ and $\beta$, using (\ref{abuv}). 
However, from Eq.\ (\ref{eqn}) the squares of
$\mbox{v}_\pm$ (or of $\alpha$ and $\beta$) are constrained by the equation of a
circle since 
\begin{equation} 
n = \langle \hat n\rangle  = \Omega (\mbox{v}_+^2 + \mbox{v}_-
^2)=2\Omega (\alpha ^2+\beta ^2).
\label{eqnv}
\end{equation} 
Thus, for a fixed particle number $n$ we may evaluate matrix
elements with only a single variational parameter, say
$\beta$, which may in turn be related to standard order parameters by comparing
matrix elements. For example,  the $z$ component of the staggered magnetization 
is related to $\beta$ and $v_\pm$ by
\begin{eqnarray} 
Q &\equiv& \langle {\cal Q}_z \rangle  = \tfrac12 \Omega
(\mbox{v}_+^2 - \mbox{v}_-^2)
\nonumber
\\ &=& 2\Omega \beta (n / (2\Omega) - \beta^2)^{1/2}. 
\label{eqq}
\end{eqnarray} 
These measures of antiferromagnetic order are in turn related 
to the superconducting order parameter $\alpha$ through Eq.\ (\ref{eqnv}). From 
Eqs.\ (\ref{eqnv}) and (\ref{eqq}), the ranges of
$\beta$ and $\alpha$ are 
$$ 0\le\beta\le\sqrt{n/4\Omega} 
\qquad
\sqrt{n/4\Omega}\le\alpha\le \sqrt{n/2\Omega}.
$$ Using Eqs.\,(\ref{eqd})--(\ref{eqq}) 
one can then evaluate the energy surface 
as a function of $Q$ or $\beta$ or $\alpha$, and study the ground state 
properties of the $SU(4)$ model. 

\section{Coherent State Energy Surfaces}

The most general Hamiltonian for the $SU(4)$ model \cite{gui99} is
\begin{eqnarray}
H = \varepsilon \hat{n} &-&\mbox{v}\ \hat{n}^2 -G_0 D^\dag D 
-G_1\vec{\pi}^\dag\cdot\vec{\pi}\nonumber \\           
&-&\chi\vec{Q}\cdot\vec{Q}+g\vec{S}\cdot\vec{S} ,         
\label{eqh1}
\end{eqnarray} 
where $\varepsilon$, v, $G_0$, $G_1$, $\chi$, and $g$ are
parameters defining the strengths of single-particle and interaction terms. 
Since
$C_{su(4)}$ is an $SU(4)$ invariant, if we assume for the ground state spin that
$\langle\vec{S}\rangle=0$ and that the number of particles
$n$ is a good quantum number, Eqs.\ (\ref{eq8})--(\ref{casimir}) imply that the
Hamiltonian (\ref{eqh1}) may be parameterized without loss of generality as
\begin{equation} 
H =H_0 -\tilde{G}_0\, [\ (1-p)D^{\dagger }D +
p\overrightarrow{{\cal Q}}\cdot \overrightarrow{{\cal Q}}\ ],
\label{eq20}
\end{equation} 
where $p$ lies in the interval 0 to 1 and
\begin{eqnarray}
G^{(0)}_{\rm{eff}}&=& (1-p)\tilde{G}_0 = G_0-G_1 \nonumber
\\ 
\chi_{\rm{eff}}&=& p\tilde{G}_0 = \chi-G_1 
\label{effcgx}  
\\ H_0 &=& \varepsilon n - \mbox{v}{n}^2 -G_1\frac{1-x^2}{4}.
\nonumber
\end{eqnarray} 
$G^{(0)}_{\rm{eff}}$ and $\chi_{\rm{eff}}$ are the effective strengths of pairing and the
$\vec{\cal Q}\cdot\vec{\cal Q}$ interactions, respectively.

From (\ref{eqnv})--(\ref{eqq}), one can show that
\begin{equation}
\mbox{v}^2_\pm=\frac{n}{2\Omega}\pm\frac{Q}{\Omega} .
\end{equation}
Eqs.\ (\ref{eqd}) and (\ref{eq13a2}) can be written as
\begin{eqnarray}
\Delta&\equiv&\langle D^{\dagger}\rangle =\langle D\rangle=\sqrt{ D^{\dagger}D}=
\Delta_+ +\Delta_- , 
\\
\Pi &\equiv&\langle \pi^{\dagger}_z\rangle =\langle \pi_z\rangle =\sqrt{
\vec{\pi}^{\dagger}\cdot\vec{\pi}}=\Delta_+ -\Delta_-
\end{eqnarray}  
where
\begin{equation}
\Delta_\pm =
\tfrac12 \Omega \sqrt{\frac14-\left(\frac{Q}{\Omega}\mp\frac x2\right)^2}
\end{equation}  
and $x$ is the effective hole concentration 
\begin{equation} 
x=1-\frac{n}{\Omega}. 
\end{equation}  
(By ``effective" we mean that
$x$ is a ratio of the hole-pair number $(\Omega-n) /2$ to  the pair degeneracy
$\Omega/2$,  rather than the  ratio of hole number to the total number of 
lattice
sites.) It can be estimated that to avoid hole-pair collapse, $\Omega$ is 
required to be roughly one-third of the total lattice sites,  and in turn the 
true hole concentration is  one-third of $x$ \cite{wu01}. 

The quantities
$\Delta$ and $\Pi$ present the spin-singlet and  spin-triplet pairing
correlations.  The former is proportional to the superconducting pairing gap; 
the latter can be regarded as a measure of the $SO(5)$ correlation since the 
$SO(5)$ Casimir operator is 
\begin{equation} 
C_{so5}=\vec{\pi}\cdot\vec{\pi}+\vec{S}\cdot\vec{S}+M(M-3) .
\end{equation}
By utilizing  Eqs.\ (\ref{eqd2})--(\ref{eqq2}), a general expression for the
energy surface of the
$SU(4)$ Hamiltonian as a function of the antiferromagnetic order parameter $Q$ 
may be obtained. In the $\Omega\rightarrow \infty$ limit, the energy surface is 
defined by
\begin{equation}
E(Q)=
\langle H\rangle-H_0=-\tilde{G}_0\, [ (1-p)\Delta^2 + P Q^2
 ] .
\end{equation} 
Converting $Q$ into the alternative order parameter 
$\beta$ allows us to express  the energy surface as a function of $\beta$ and 
$n$,
\begin{eqnarray} 
&&E(\beta)=\langle H\rangle-H_0=-\frac{\tilde{G}_0\Omega^2}4
\nonumber
\\ &&\times\left\{ (24p-8)\beta^2\left(\frac{n}{2\Omega}-\beta^2\right)+2(1-p)
\left[\
\frac{n}{2\Omega}\left(1-\frac{n}{2\Omega}\right)\nonumber
\vphantom{\sqrt{ \left( \left( \frac{n}{2\Omega} \right) ^ 2 \right) } }
\right.\right.
\\ &&+\left.\left.\left(
\frac{n}{2\Omega}-2\beta^2
\right)
\sqrt{
\left(1-\frac{n}{2\Omega}\right)^2-4\beta^2\left(\frac{n}{2\Omega}-
\beta^2\right)}\
\right]\right\}
\label{eqeb}
\end{eqnarray} 
which may also be expressed in terms of the  superconducting 
order parameter $\alpha$ using (\ref{eqnv}).

\section{Coherent States and SU(4) Subgroups}

Assuming $\tilde{G}_0 > 0$ (suggested by phenomenology), 
$p = 1/2$ in Eq.\ (\ref{eq20}) corresponds to
$SO(5)$ symmetry, while the extreme values 0 and 1 correspond to  $SU(2)$ and
$SO(4)$ symmetries, respectively (see Ref.\
\cite{gui99}). Other values of $p$ respect $SU(4)$ symmetry but break the 
$SO(5)$, $SO(4)$, and $SU(2)$ subgroups.   In Fig.\ 1 we illustrate the
ground-state energy $E(\beta)$ of Eq. (\ref{eqeb}) 
as a function of the  order parameter
$\beta$ for different electron occupation fractions $n/\Omega$ with $p = 0, 
{\frac{1}{2}}$ and 1. 

\epsfigwide{fig1}{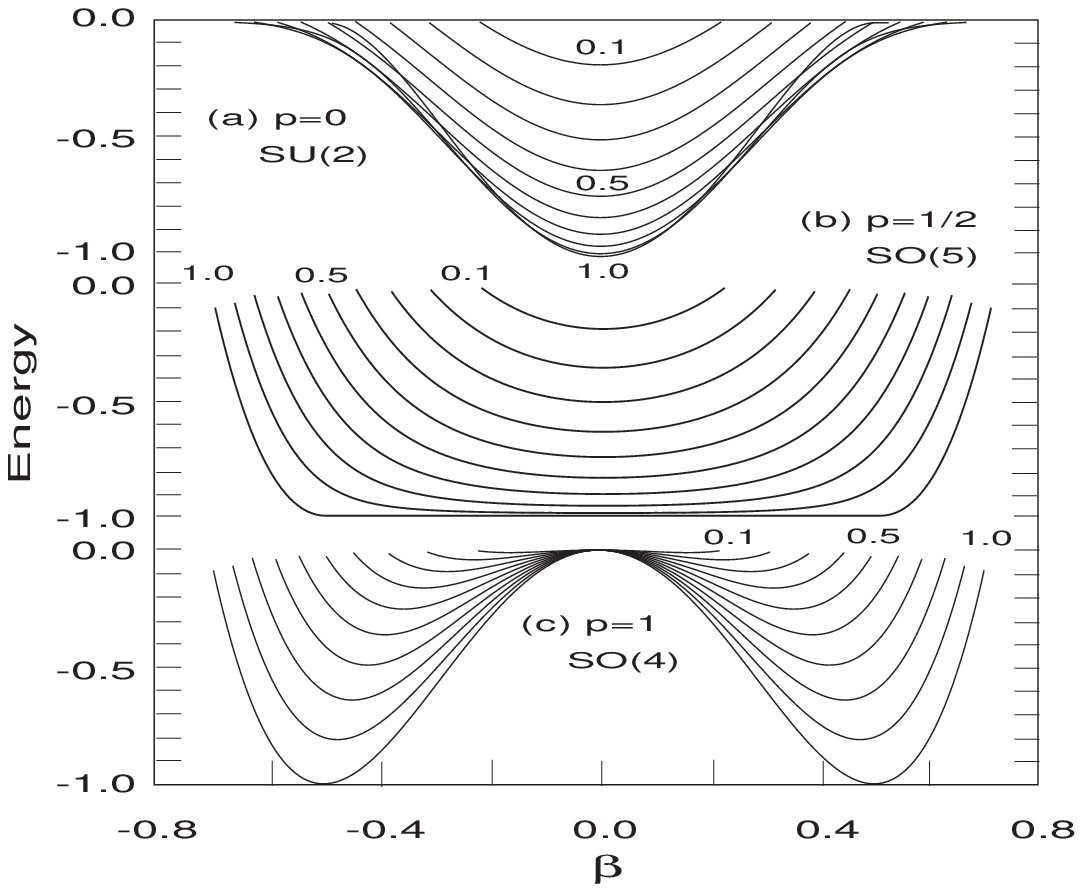}{0pt}{0pt }
{Coherent state  energy surfaces.  The energy units are 
$G^{(0)}_{\rm{eff}} \Omega^2 / 4$ for figures (a) and (b), and $\chi_{\rm{eff}}
\Omega^2 / 4$ for (c). 
$H_0$ is taken as the energy zero point.  Numbers on curves are  the
effective lattice occupation fractions,  with $n / \Omega = 1$ corresponding to 
half
filling and $0 < n / \Omega < 1$ to finite hole doping.
$SO(5)$ symmetry corresponds to $p=1/2$ and  the allowed range of $\beta$ is
$[-\frac12
\sqrt{n/\Omega},\frac12\sqrt{n/\Omega}]$, which depends on $n$. The order
parameter $\beta$ is related to the order parameter $Q \equiv \langle {\cal Q}_z
\rangle$ (staggered magnetization)  and the electron number $n$ through Eq.\
(\ref{eqq}).}

\subsection{SU(2) Limit}

For $p = 0$ [$SU(2)$ limit; see Fig.\ 1a], 
the minimum energy occurs at $\beta = 0$ 
(equivalently, $Q=0$) for all values of $n$. Thus, $\Delta$ reaches its maximum
value of
\begin{equation} 
\Delta_{\rm max} = \tfrac12 \Omega \sqrt{1-x^2}, 
\end{equation} 
indicating superconducting order.

\subsection{SO(4) Limit}

For $p = 1$ [$SO(4)$ limit; see Fig 1c], the opposite situation occurs:
$\beta = 0$ is an unstable point and an infinitesimal fluctuation will drive the
system to the energy minima at finite, 
\begin{equation}
\beta = \pm \tfrac12\sqrt{n/\Omega}. 
\end{equation} 
Thus, $|Q|$ reaches its maximum value of $n/2$, indicating the
presence of AF order.

\subsection{SO(5) Limit}

From Fig.\ 1b, the
$SO(5)$ dynamical symmetry (with $p={\frac {1}{2}}$) 
is seen to have extremely interesting behavior: the
minimum energy occurs at $\beta = 0$ for all values of $n$, as in the
$SU(2)$ case, but there are large-amplitude  fluctuations in AF and SC order. In
particular, when $n$ is near $\Omega$ (half filling), the system has an energy
surface almost flat for broad ranges of $\beta$ (or $Q$ or $\alpha$). This 
suggests a phase having much of the character of a ``spin glass'' for a range of
doping fractions (specifically, a phase very soft against fluctuations in the order
parameters). It could also lead to inhomogeneous structures such as stripes if
there is a periodic spatial modulation of the system, since the soft nature of 
the energy surface implies that relatively small perturbations can shift an $SO(5)$
system between antiferromagnetic and superconducting behavior. As $n/\Omega$
decreases, the fluctuations become smaller and the energy surface tends more and
more to the $SU(2)$ (superconducting) limit.

\subsection{SO(5) As a Critical Dynamical Symmetry}

Dynamical symmetries that, within the dynamical symmetry itself,
exhibit a transition between qualitatively
different energy surfaces as a parameter (usually related to particle
number) is varied have been termed {\em critical dynamical symmetries}
\cite{wmzha87}. The
$SO(5)$ dynamical symmetry, within the context of its $SU(4)$ parent symmetry,
exhibits such
transitional properties.
At half filling the energy surface is completely flat under variations
of the antiferromagnetic
order parameter $\beta$ (see the $n=1.0$ curve of Fig.\ 1b), implying
large fluctuations in the order parameters. 
But as hole doping is increased 
the $SO(5)$ energy surface changes smoothly into one localized around
$\beta = 0$ (see the $n=0.1$ curve of Fig.\ 1b).  
Thus, $SO(5)$ is an example of a critical dynamical
symmetry.

Such symmetries are well known in nuclear
structure physics \cite{wmzha87,wmzha88a,wmzha90}. 
The $SO(5)$ critical dynamical
symmetry discussed here in a condensed matter context has many formal similarities
with the $SO(8)\supset SO(7)$ critical dynamical symmetry of the (nuclear) Fermion
Dynamical Symmetry Model \cite{wu94}.  
The condition for realization of
the $SO(5)$ critical dynamical symmetry is that the strength of
${\cal Q}\cdot {\cal Q}$ equals that of $D^{\dag} D$ in the Hamiltonian; This is
similar to the $SO(7)$ nuclear critical dynamical symmetry, which is realized 
when there is an overall $SO(8)$ symmetry and the monopole pairing and the 
quadrupole interaction terms are of equal strength in the nuclear Hamiltonian 
\cite{wmzha87}.
In the $SO(7)$ case of nuclear physics, the order parameter analogous to $\beta$
presents nuclear deformation: nuclei
around midshell (half filling of a shell by nucleons) are soft 
against shape fluctuations and 
transform into a spherical shape (the $SO(5)\times SU(2)$ dynamical
symmetry limit of the $SO(8)$ symmetry) 
as the number of nucleons increases.

\section{SO(5) Symmetry Breaking}

Under exact
$SO(5)$ symmetry ($p={\frac {1}{2}}$),
the AF and SC states are degenerate at half filling. There is
no barrier between AF and SC states,
and one can fluctuate into the other at zero
cost in energy (see the $n/\Omega =1$ curve of Fig.\ 1b).
This situation is inconsistent with Mott insulating behavior at
half-filling.  The Zhang $SO(5)$ model has been challenged because
under exact symmetry it does not fully respect
the phenomenological requirements of ``Mottness".
As Zhang \cite{zha97} has recognized, for antiferromagnetic insulator properties
 to exist at half filling,
it is necessary to break
$SO(5)$ symmetry.  Such breaking of the $SO(5)$ subgroup symmetry is implicit in
 the $SU(4)$ model, occurring naturally
in the $SU(4)$ model if $p > 1/2$ in the Hamiltonian
(\ref{eq20}).  Furthermore, the $SU(4)$ symmetry leads to
the following constraint
\begin{equation}
\langle D^\dagger D+\vec{\cal Q}\cdot \vec{\cal
Q}+\vec{\pi}\cdot\vec{\pi}\rangle
=\frac{1-x^2}{4}\Omega^2 .
\end{equation}
This ensures a doping dependence in the solutions,
which is necessary for describing the transition from AF to SC in the cuprates.

Thus, the coherent state analysis indicates that the phenomenologically required
$SO(5)$ symmetry breaking and the doping dependence in the solutions occur
naturally in the $SU(4)$ model.
They need not be introduced empirically as proposed in the original Zhang 
$SO(5)$ model \cite{zha97}.
Recently, a projected $SO(5)$ model
\cite{zha99} has been introduced. Its essence is a
patch to the original $SO(5)$
model that implements the Gutzwiller projection in
order to satisfy the large-$U$ Hubbard (non-double-occupancy or the
Mott-insulator) constraint.  In our $SU(4)$ model, there is no need
to introduce such a projection artificially
because the $SU(4)$ symmetry constraint {\em already implies} a
constraint of non-double-occupancy with charge density localized on sites of the
underlying lattice.  We shall demonstrate this explicitly and give a detailed 
discussion of the consequences in a subsequent paper.

To see in more detail how in the $SU(4)$ model a broken $SO(5)$ symmetry can
interpolate between AF and SC states as particle number varies,  let us perturb
slightly away from the $SO(5)$ limit of $p = 1/2$ in Eq.\ (\ref{eq20}).  In 
Fig.\
2a, $SU(4)$ coherent state results for $p = 0.52$ are shown. 
We denote the value 
of $\beta$ minimizing $\ev H$ as
$\beta_0$. The corresponding variation of the AF order parameter
$Q=\langle{\cal Q}_z\rangle$ with $n$, and its comparison with 
the variation in various
symmetry limits are summarized in Fig.\ 2b. The variations of the AF, SC and 
SO(5)
correlations ($Q$, $\Delta$  and $\Pi$) with the hole doping $x$ are shown in 
Fig.\ 2c, while the variations of the contributions of each term in the Hamiltonian
(the pairing $D^\dagger D$ and the AF interaction 
$\vec{\cal Q}\cdot \vec{\cal Q}$)
to the total energy are shown in Fig.\ 2d.
In Fig. 2, there is an important quantity, the critical doping $x_c$,
which can be expressed analytically as \cite{wu01}
\begin{equation}
x_c=\sqrt{1-\frac{G^{(0)}_{\rm eff}}{\chi_{\rm
eff}}}=\sqrt{1-\frac{1-p}{p}} .
\label{cdoping}
\end{equation}
For $p=0.52$, we have $x_c=0.277$. 

\epsfigwide{fig2}{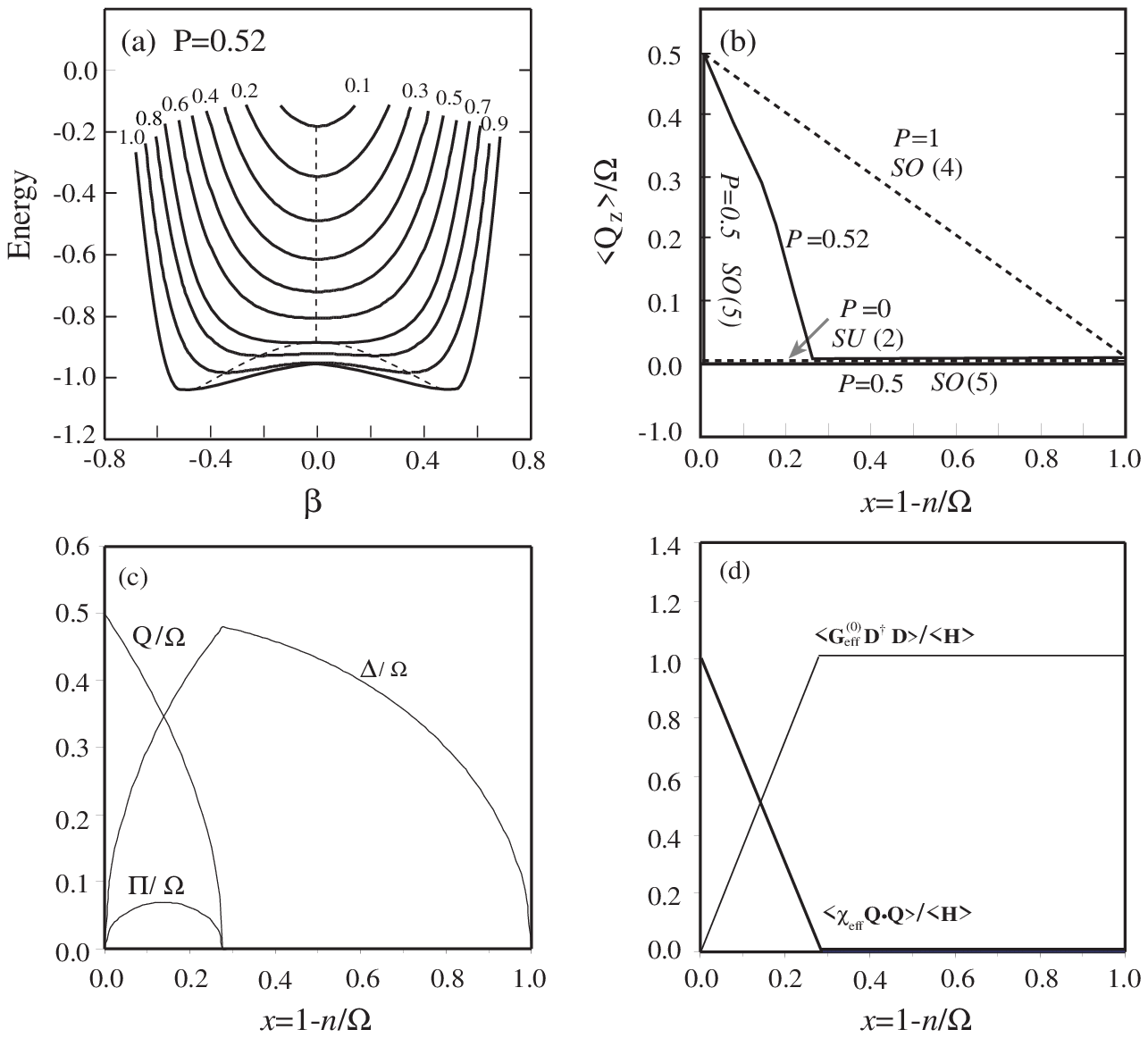}{0pt}{0pt }
{(a)~As for Fig. 1, but for slightly perturbed $SO(5)$. The dotted line
indicates the location of the ground state as $n$ varies
and the corresponding $\beta$ value is denoted as $\beta_0$. 
(b)~Variation of the 
AF order parameter with effective occupation number for different values of 
$p$. (c)~Variation of the AF, SC and $SO(5)$ order parameters $Q$, 
$\Delta$, and
$\Pi$ as functions of the effective hole concentration $x$. (d)~Variation of the
ratio of pairing and $\vec{\cal Q}\cdot\vec{\cal Q}$ interactions to the total
energy of the system  as functions of the effective hole concentration $x$. }
 
\subsection{Antiferromagnetic Order}
 
One sees from Fig.\ 2 that if $n$ is near $\Omega$
(half filling), $\beta_0 \simeq \pm 0.5$;  this corresponds to AF order, since 
the staggered magnetization reaches
its maximum, $Q=\Omega/2$, and there are no pairing
or $SO(5)$ correlations ($\Delta=\Pi=0$).   With the onset of hole doping,
$n/\Omega$ decreases ($x$ increases).  The AF correlation $Q$ quickly diminishes
and the pairing and $SO(5)$  correlations, $\Delta$ and $\Pi$, increase. 

\subsection{Underdoped SO(5) Fluctuations}

Before $Q$ vanishes at the critical doping 
$x_c=0.277$ ($n/\Omega=0.723$), 
the system has an energy surface almost flat for 
broad ranges of $\beta$, implying the presence of large-amplitude fluctuations 
in AF order (and equivalently in SC order).  Meanwhile, the $SO(5)$ correlation 
$\Pi$ increases and reaches its maximum at the doping where the pairing and AF
correlations become equal to each other (see Figs.\ 2c and 2d). This is 
the underdoped SC region. The coexistence of these three correlations
competing with each other is consistent with the complexity and variety of
experimental phenomena in this region. 

\subsection{Superconducting Order}

For small values of $n$ ($x>x_c$), the stable point is $\beta_0 = 0$. This
corresponds to SC order, since both the AF and $SO(5)$ correlations vanish
($Q=\Pi=0$), and only the pairing correlation remains ($\Delta >0$). The
critical doping $x_c$ is the optimal doping point since 
$\Delta$ is maximum at $x_c$  and decreases as hole-doping increases.  Thus the
doping range $x>x_c$ may be considered to be the overdoped region. The critical
doping $x_c$  depends on the ratio of the pairing and the
$\vec{\cal  Q}\cdot
\vec{\cal Q}$ strengths (see Eq. (\ref{cdoping})). 
The larger the pairing strength $ G^{(0)}_{\rm eff}$ relative to 
the $\vec{\cal Q}\cdot \vec{\cal Q}$ strength $\chi_{\rm eff}$ in Eq.
(\ref{cdoping}),  the smaller the
critical doping value $x_c$.

\section{Hamiltonian for High T$_c$ Superconductivity}

From the above discussion one can see that, with a perturbed $SO(5)$ symmetry, 
the system can undergo phase transitions from the AF order
at half filling to the SC order at smaller filling 
as particle number varies. 
This picture is at least qualitatively
consistent with the observations. The $SO(5)$ symmetry breaking in the
Hamiltonian ($p$ larger than $1/2$) is crucial. 
Only when $SO(5)$ is broken does the  energy surface 
interpolate between AF and SC order as doping is varied (compare the surfaces for
$p=1/2$  and $p=0.52$ in Figs.\ 1 and 2). 
We thus conclude that 
high temperature superconductivity may be described by a Hamiltonian that
conserves $SU(4)$ but breaks (explicitly) 
$SO(5)$ symmetry in a direction favoring AF order over SC order. 

The deviation 
from the $SO(5)$ symmetry need not be large. Experimentally, it
is known that the optimal doping $P_c$ is around 0.16, 
suggesting that $x_c\simeq 0.48$
(note that $x_c\simeq 3P_c$). This leads to $p=0.56$
according to Eq.\ (\ref{cdoping}), which is formally quite close to the
$SO(5)$ symmetry limit. 

However, it should be stressed that because of the critical nature of the $SO(5)$
symmetry, a slightly perturbed $SO(5)$ Hamiltonian may have solutions that are
close to the other symmetry limits of the
$SU(4)$ model for particular electron  occupation ratios, even though the
Hamiltonian itself is not formally  in any of the dynamical symmetry limits.  For
example, one can see from Figs.\ 2c  and 2d that the Hamiltonian near half filling
actually behaves like an $SO(4)$ Hamiltonian since it effectively contains only the
$\vec{\cal Q}\cdot\vec{\cal Q}$ correlation term ($\vec{\cal Q}\cdot\vec{\cal Q}$
is the primary component of  the $SO(4)$ Casimir). Likewise the perturbed $SO(5)$
Hamiltonian approximates the $SU(2)$  Hamiltonian containing only the $D^\dagger D$
term when $x>x_c$ (see Figs 2c and 2d) .
Only in the intermediate doping range ($0<x<x_c$), where both $D^\dagger D$ and 
$\vec{\cal Q}\cdot\vec{\cal Q}$ correlations have significant contributions,  
does the $p=0.52$ Hamiltonian behave like an approximate $SO(5)$  Hamiltonian, as it
should formally.  In particular, near the region  where the $D^\dagger D$ and
$\vec{\cal Q}\cdot\vec{\cal Q}$ terms have equivalent contributions and the 
$SO(5)$ correlations (the operator $\Pi$) reach their maximum, the $p=0.52$ 
solution  lies very close to the $SO(5)$ symmetry limit, as one would expect.

The present analysis implies that the underdoped
regime is naturally associated with the $SU(4) \supset SO(5)$ dynamical 
symmetry interpolating between antiferromagnetic and superconducting order.
Likewise, optimally doped and overdoped superconductors are naturally associated
with the
$SU(4) \supset SU(2)$ dynamical symmetry and AF insulators near half filling are
associated with the  
$SU(4) \supset SO(4)$ dynamical symmetry (or small perturbations around these
symmetries).  

As we shall discuss in a separate publication \cite{sun99},  the appearance of
pseudogap behavior \cite{pseudogap} can be described,  and much of the 
quantitative phase diagram in cuprates can be reproduced rather well \cite{wu01},
if a fixed Hamiltonian with slightly broken $SO(5)$ symmetry but preserving
$SU(4)$ overall symmetry is adopted.  This again supports our interpretation of
$SO(5)$ symmetry as an critical dynamical symmetry. The small $SO(5)$ symmetry
breaking distorts the completely flat energy surface at half filling
to stabilize AF character in the system; the
critical nature of the $SO(5)$ dynamical symmetry (that the system interpolates
between two phases) remains. 

\section{Summary and Conclusions}

The present paper serves to introduce into the condensed matter literature the
technology of generalized coherent states.   As we have shown in the example
discussed here, these  methods provide a systematic way to relate a many-body
theory to its approximate broken symmetry solutions.  This approach may be 
viewed as a standardized technology for constructing energy surfaces for many-body 
theories defined in terms of the algebra of their second-quantized operators, or
equivalently as the most general Bogoliubov transformation permitted, 
subject to a symmetry constraint on the Hamiltonian of a system.

To illustrate the power of these techniques,  we have used generalized coherent
states  to understand the relationship between the $SU(4)$ model of
superconductivity \cite{gui99} and the Zhang $SO(5)$ model \cite{zha97}. The use of
$SU(4)$  coherent states to analyze the energy surface of its $SO(5)$ subgroup
permits us to interpret the $SO(5)$ as a critical dynamical symmetry that
interpolates between AF and $d$-wave SC order as doping is varied, and suggests 
similarities with analogous critical dynamical symmetries well known from nuclear
structure physics.  This permits the $SO(5)$ symmetry to be understood dynamically
as a critical phase that, for a  range of doping, has an energy surface extremely
soft against AF fluctuations and therefore having much of the character of a spin
glass (or possible stripe  phases in the presence of a spatially modulated
perturbation).   

Thus, the coherent  state analysis suggests
that $SO(5)$ is the appropriate symmetry of the underdoped regime, but that 
the AF phase at half filling and the optimal and overdoped SC phases are described 
by two other $SU(4)$ subgroups:  $SO(4)$ and $SU(2)$, respectively.  The coherent
state analysis also shows clearly that the requirement of 
small deviations from $SO(5)$ symmetry
and the necessary doping dependence of the solutions that are  
inserted in the Zhang model occur naturally when
$SO(5)$ is a subgroup of $SU(4)$. 

In addition, we note that the results obtained here may have some broader
implications.  Although the present application is specifically to the
high-temperature superconductor problem, we may anticipate that these 
mathematical 
techniques could find use for any application in condensed matter physics where 
it is important to understand the relationship between an exact many-body 
theory 
and the order parameter(s) characterizing its approximate broken symmetry
solutions.  Clearly there are many such possibilities.

Finally, the concept of a critical dynamical symmetry that we have 
introduced
here in a condensed matter context is one that has already found important
application in other areas of physics. This implies that there may be deep 
algebraic
analogies between various condensed matter systems and superficially different
systems appearing in other fields of many-body physics.  We have suggested one 
such analogy here between the physics of high-temperature superconductors and 
the physics of collective states in heavy atomic nuclei.

L.-A. Wu was supported in part by the National Natural Science Foundation of 
China.

\baselineskip = 14pt
\bibliographystyle{unsrt}

\begin{thebibliography} {99}
\bibitem{note1} The $SU(4)$ model employs 16 generators closed under commutation
that define the $U(4)$ Lie algebra and a corresponding $U(4)$ group. 
In addition
to the three components of staggered magnetization and two for $d$-wave pairing,
there are three spin operators, one charge operator, six spin-1 operators
analogous to the Zhang $\pi$ operators, 
and one operator that may be interpreted 
as generating charge density waves.  The $U(4)$ group has the subgroup $U(4) 
\supset
U(1) \times SU(4)$, where the $U(1)$ factor may be identified with charge 
density
waves.  Thus, because of the direct product structure, one can without loss of
generality view the theory as an $SU(4)$ theory describing superconductivity and
antiferromagnetism, and global charge and spin conservation, with the $U(1)$
charge-density wave sector treated independently (but the overall group 
structure
implies that the lowest states of the symmetry-limit theory have no charge-
density
wave excitations).
\bibitem{CDwave} Charge density waves are forbidden in the unbroken-pair 
subspace
with  exact $SU(4)$ symmetry.  A relaxation of the exact symmetry or breaking 
pairs
can lead to charge density wave excitations in a more realistic theory.  
However,
the clear tendency of the $SU(4)$ symmetry is to suppress charge density waves 
in
the lowest-lying states.
\end{thebibliography}

\end{document}